\documentclass[conference]{IEEEtran}
\IEEEoverridecommandlockouts
\usepackage[left=0.65in,right=0.65in,bottom=0.98in,top=0.72in]{geometry}
\usepackage{amsmath,amssymb,amsfonts}
\usepackage{algorithmic}
\usepackage{algorithm}
\usepackage{xcolor}
\usepackage{array}
\usepackage{subfigure}
\usepackage{color} 
\usepackage{setspace}
\usepackage{textcomp}
\usepackage{stfloats}
\usepackage{url}
\usepackage{verbatim}
\usepackage{graphicx}
\usepackage[noadjust]{cite}

\usepackage{balance}
\usepackage[colorlinks=true,linkcolor=red,citecolor=green]{hyperref}

\begin{document}
	\title{Energy Efficiency Maximization for Movable Antenna-Enhanced System Based on Statistical CSI}
	\author{\IEEEauthorblockN{Xintai Chen, Biqian Feng, Yongpeng Wu, and Wenjun Zhang}
		\thanks{X. Chen, B. Feng, Y. Wu, and W. Zhang are with the Department of Electronic Engineering, Shanghai Jiao Tong University, Minhang 200240, China (e-mail: chenxintai@sjtu.edu.cn; fengbiqian@sjtu.edu.cn; yongpeng.wu@sjtu.edu.cn; zhangwenjun@sjtu.edu.cn).}
		\thanks{Corresponding author: Yongpeng Wu.}}
	
	\maketitle
	
	\begin{abstract}
		This paper investigates an innovative movable antenna (MA)-enhanced multiple-input multiple-output (MIMO) system designed to enhance communication performance. We aim to maximize the energy efficiency (EE) under statistical channel state information (S-CSI) through a joint optimization of the transmit covariance matrix and the antenna position vectors (APVs). To solve the stochastic problem, we consider the large number of antennas scenario and resort to deterministic equivalent (DE) technology to reformulate the system EE w.r.t. the transmit variables, i.e., the transmit covariance matrix and APV, and the receive variables, i.e., the receive APV, respectively. Then, we propose an alternative optimization (AO) algorithm to update the transmit variables and the receive variables to maximize the system EE, respectively. Our numerical results reveal that, the proposed MA-enhanced system can significantly improve EE compared to several benchmark schemes and the optimal performance can be achieved with a finite size of movement regions for MAs.
	\end{abstract}
	\begin{IEEEkeywords}
		Movable antenna, energy efficiency, S-CSI.
	\end{IEEEkeywords}
	
	\section{Introduction}
	In recent years, there has been a significant surge in the demand for wireless data services, driven by the proliferation of mobile devices and cutting-edge applications such as virtual reality, cloud-based services, and edge computing. Consequently, these developments present novel challenges for the evolution of future wireless communication systems. Advanced movable antenna (MA) technology, also known as fluid antenna \cite{BLIF,CFAS}, is a promising solution to tackle these issues, as they can fully exploit the spatial variations of the wireless channel across the entire transmit/receive region by dynamically reconfiguring the physical placement of antennas \cite{Mapa}.
	
	To unlock their potential, the maximum achievable rate of point-to-point multiple-input multiple-output (MIMO) system is investigated in reference \cite{Mccf}, where the transmit and receive antenna position vectors (APVs) and the covariance matrix of the transmit signal are jointly optimized. In reference \cite{MAAM}, the discrete positions of the MAs are optimized through a greedy search-based algorithm, aiming to enhance the achievable rate of a two-user multicast multiple-input single-output system. To achieve low-complexity design, reference \cite{Biqian} proposes a planar movement mode, where each MA is only allowed to move within a given planar area, and the minimum distance between any two areas is set to avoid the coupling effect. Furthermore, the six-dimensional MA (6DMA) system is introduced in \cite{6MAB,6MAE}, where the three-dimensional (3D) positions and 3D rotations of distributed antenna surfaces can be adjusted based on the users' spatial distribution. However, most research efforts focus on designing the positions of MAs based on instantaneous channel state information (I-CSI), which is hard to obtain due to the associated high delays and power consumption. Indeed, these delays arise from various processes such as CSI estimation and/or feedback and advanced signal processing for APV design as well as physical antenna movement. The I-CSI can quickly become outdated if these delays exceed the channel coherence time.
	
	To alleviate this problem, reference \cite{JBaA} adopts a constrained stochastic successive convex approximation algorithm to maximize the average achievable rate of the point-to-point MIMO system exploiting statistical channel state information (S-CSI) by jointly optimizing the APVs and the transmit covariance matrix. In addition, references \cite{S-CSI LiYou, JBaAEE} utilize Jensen's inequality to simplify the achievable rate with its tight upper-bound deterministic function, and then apply the alternating optimization (AO) method to solve each subproblem. In the future, as the base station (BS) tends to accommodate a massive number of antennas, the size, power consumption, hardware cost, and computational complexity of the transceivers will be largely increased.
	
	In this paper, we aim to exploit S-CSI to maximize energy efficiency (EE) for MA-enhanced MIMO systems. The main contributions of this paper can be summarized as follows:
	\begin{itemize}
		\item [1)] Firstly, We investigate the average EE maximization for large-scale MA-enhanced MIMO systems based on S-CSI. To handle the difficulty of expectation, we leverage the deterministic equivalent (DE) technique to reformulate the average achievable rate with respect to the transmit variables, i.e., the transmit covariance matrix and APV, and the receive variables, i.e., the receive APV, respectively.
		
		\item [2)] Secondly, we develop an AO algorithm to maximize the system EE by iteratively optimizing the transmit variables and the receive variables based on the reformulated deterministic objective functions. In each subproblem, we efficiently develop a stationary point with the successive convex approximation (SCA) method.
		
		\item [3)] Finally, we provide extensive numerical results to reveal the significant gains in system EE realized with the proposed MA-enhanced MIMO system compared to several benchmarks, including those employing conventional fixed-position uniform planar arrays (UPAs) at the transmitter (Tx) and/or the receiver (Rx) as well as MAs with discrete movement at the transceiver.
	\end{itemize}
	
	The remainder of this paper is organized as follows. Section 
	\ref{sec: system model} presents the considered MA-enhanced MIMO model based on S-CSI and EE maximization problem. Section \ref{sec: algorithm} adopts DE algorithm to approximate the expectation function, and adopts AO algorithm to solve it. Numerical results and corresponding discussions are presented in Section \ref{sec: numerical results}. Finally, Section \ref{sec: conclusion} concludes this paper.
	
	\textbf{Notations}: $(\cdot)^{T}$ and $(\cdot)^{H}$ denote the transpose and conjugate transpose (Hermitian), respectively. $\odot$ denotes the Hadamard product. $\left[ \mathbf{x} \right]_{\mathcal{X}}$ projects $\mathbf{x}$ into region $\mathcal{X}$. $\left[ \mathbf{X} \right]^+$ projects the elements of $\mathbf{X}$ into the non-negative real number field. $[\mathbf{A}]_{p_1,\cdots,p_K}$ denotes the entry with index $p_1,\cdots,p_K$ of $K$-dimension matrix $\mathbf{A}$. 
	
	\section{System Model and Problem Formulation}
	\label{sec: system model}
	We consider a single-user MIMO communication system, where Tx and Rx are equipped with $N$ and $M$ MAs, respectively. Assume that the transmit MAs and the receive MAs can move within square regions $\mathcal{C}_{\mathrm{t}}$ and $\mathcal{C}_{\mathrm{r}}$, respectively. That is, the position of transmit antenna $n$ satisfies $\mathbf{t}_n \triangleq \left(x_{\mathrm{t},n},y_{\mathrm{t},n}\right) \in \mathcal{C}_{\mathrm{t}}$ and the position of receive antenna $m$ satisfies $\mathbf{r}_{m} \triangleq \left(x_{\mathrm{r},m},y_{\mathrm{r},m}\right) \in \mathcal{C}_{\mathrm{r}}$. By stacking the coordinates of all MAs together, the transmit and receive APVs are denoted by $\mathbf{t}=\left(\mathbf{t}_1, \mathbf{t}_2, \cdots, \mathbf{t}_N\right)^{T}$ and $\mathbf{r}=\left(\mathbf{r}_{1}, \mathbf{r}_{2}, \cdots, \mathbf{r}_{M}\right)^{T}$, respectively. To avoid potential coupling, a minimum distance $D \geq \lambda/2$ is required between each pair of MAs \cite{AAfL}, i.e., $\left\|\mathbf{t}_i-\mathbf{t}_j\right\| \geq D, \forall i \neq j$, and $\left\|\mathbf{r}_{i}-\mathbf{r}_{j}\right\| \geq D$, $\forall i \neq j$, where $\lambda$ is the wavelength of the signal carrier. Then, the received signal is given by
	\begin{equation}
		\mathbf{y}=\mathbf{H}(\mathbf{t},\mathbf{r}) \mathbf{x}+\mathbf{z}.
	\end{equation}
	Here, $\mathbf{H}\left(\mathbf{t},\mathbf{r}\right) \in \mathbb{C}^{M \times N}$ represents the channel matrix between Tx and Rx, which is related to APVs $\mathbf{t}$ and $\mathbf{r}$. The transmitted signal, $\mathbf{x}$, follows a zero-mean CSCG distribution with covariance matrix $\mathbb{E}\left\{\mathbf{x} \mathbf{x}^{H}\right\}=\mathbf{Q}$. $\mathbf{z} \sim \mathcal{CN}\left(\mathbf{0}, \sigma^2 \mathbf{I}_{M}\right)$ denotes the CSCG noise with zero mean and variance $\sigma^2$.
	
	\subsection{Channel Model}
	We assume that Rx is in the far field of Tx, and thus, the angles of departure (AoDs) and angles of arrival (AoAs) for different positions in regions $\mathcal{C}_{\mathrm{t}}$ and $\mathcal{C}_{\mathrm{r}}$ are identical, respectively \cite{Mapa,Mccf}. We assume the Rician channel model holds and there are $L_{\mathrm{t}}$ transmit paths and $L_{\mathrm{r}}$ receive paths overall. Let $(\theta_{\mathrm{t}}^{l_{\mathrm{t}}}, \phi_{\mathrm{t}}^{l_{\mathrm{t}}})$ and $(\theta_{\mathrm{r}}^{l_{\mathrm{r}}}, \phi_{\mathrm{r}}^{l_{\mathrm{r}}})$ denote the elevation and azimuth angles of the $l_{\mathrm{t}}$-th transmit path and the $l_{\mathrm{r}}$-th receive path, respectively. Then, the difference in the signal propagation distance between position $\mathbf{t}_n$ and the origin of $\mathcal{C}_{\mathrm{t}}$ and that between position $\mathbf{r}_{m}$ and the origin of $\mathcal{C}_{\mathrm{r}}$ are, respectively, given by
	\begin{equation}
		\begin{aligned}
			\rho_{\mathrm{t}}^{l_{\mathrm{t}}}(\mathbf{t}_n)&=x_{\mathrm{t},n} \sin \theta_{\mathrm{t}}^{l_{\mathrm{t}}} \cos \phi_{\mathrm{t}}^{l_{\mathrm{t}}}+y_{\mathrm{t},n} \cos \theta_{\mathrm{t}}^{l_{\mathrm{t}}},\\
			\rho_{\mathrm{r}}^{l_{\mathrm{r}}}(\mathbf{r}_{m})&=x_{\mathrm{r},m} \sin \theta_{\mathrm{r}}^{l_{\mathrm{r}}} \cos \phi_{\mathrm{r}}^{l_{\mathrm{r}}}+y_{\mathrm{r},m} \cos \theta_{\mathrm{r}}^{l_{\mathrm{r}}}.
		\end{aligned}
	\end{equation}
	Then, the field response vectors of MA $n$ at Tx and MA $m$ at Rx are, respectively, given by
	\begin{equation}
		\begin{aligned}
			\mathbf{g}(\mathbf{t}_n) \triangleq & \left[e^{j \frac{2 \pi}{\lambda} \rho_{\mathrm{t}}^{1}(\mathbf{t}_n)}, \cdots, e^{j \frac{2 \pi}{\lambda} \rho_{\mathrm{t}}^{L_{\mathrm{t}}}(\mathbf{t}_n)}\right]^{T},\\
			\mathbf{f}(\mathbf{r}_{m}) \triangleq & \left[e^{j \frac{2 \pi}{\lambda} \rho_{\mathrm{r}}^{1}(\mathbf{r}_{m})},\cdots, e^{j \frac{2 \pi}{\lambda} \rho_{\mathrm{r}}^{L_{\mathrm{r}}}(\mathbf{r}_{m})}\right]^{T}.
		\end{aligned} 
	\end{equation}
	Furthermore, the channel matrix $\mathbf{H}(\mathbf{t},\mathbf{r})$ can be written as
	\begin{equation}
		\label{eq: discrete channel matrix}
		\begin{aligned}	
			\mathbf{H}(\mathbf{t},\mathbf{r})
			= \mathbf{F}^H\left(\mathbf{r}\right) \mathbf{\Sigma} \mathbf{G}\left(\mathbf{t}\right),
		\end{aligned}
	\end{equation}
	where $\mathbf{F}(\mathbf{r}) \triangleq\left[\mathbf{f}\left(\mathbf{r}_{1}\right), \cdots, \mathbf{f}\left(\mathbf{r}_{M}\right)\right] \in \mathbb{C}^{L_{\mathrm{r}} \times M}$ and $\mathbf{G}(\mathbf{t}) \triangleq\left[\mathbf{g}\left(\mathbf{t}_1\right), \cdots, \mathbf{g}\left(\mathbf{t}_N\right)\right]\in \mathbb{C}^{L_{\mathrm{t}} \times N}$ denote the field response matrices at Tx and Rx, respectively. $\mathbf{\Sigma}\in\mathbb C^{L_{\mathrm{r}}\times L_{\mathrm{t}}}$ is the path response matrix, which represents the gains of all channel paths from the origin of $\mathcal{C}_{\mathrm{t}}$ to the origin of $\mathcal{C}_{\mathrm{r}}$. 
	
	In MA-enhanced systems, the acquisition of I-CSI is usually impractically costly due to the dynamic changes in the physical environment. Consequently, this paper employs the relatively stable S-CSI to design the APVs. Without loss of generality, we assume that the LOS path is in the $1$-st transmit and $1$-st receive path. Thus, the path response matrix can be given by
	\begin{equation}
		\begin{aligned}
			\mathbf{\Sigma} = \overline{\mathbf{\Sigma}} + \widetilde{\mathbf{\Sigma}}.
		\end{aligned}
	\end{equation}
	Here, matrix $\overline{\mathbf{\Sigma}}$ has zero entries except one non-negative entry in the $1$-st column and the $1$-st row. Besides, matrix $\widetilde{\mathbf{\Sigma}}$ can be written as
	\begin{equation}
		\begin{aligned}
			\widetilde{\mathbf{\Sigma}} = \mathbf{M} \odot  \mathbf{W},
		\end{aligned}
	\end{equation}
	where the entries of $\mathbf{M}$ are non-negative and represent the average path gain and the entries of $\mathbf{W}$ are independently and identically distributed (i.i.d.) zero-mean complex Gaussian random variables with unit variance. 
	
	\subsection{Problem Formulation}
	We aim to maximize the system EE utilizing the S-CSI by jointly optimizing the APVs and the transmit covariance matrix under the position constraints and the power constraint. Since Tx has access to only S-CSI, we focus on the average achievable rate, which can be expressed as \cite{Rtfm}, \cite{Bdst} 
	\begin{equation}
		\label{eq: Rk}
		\begin{aligned}
			R\left(\mathbf{t},\mathbf{r},\mathbf{Q}\right) = & \mathbb{E}_{\widetilde{\mathbf{\Sigma}}}\left\{\log 	\operatorname{det}\left(\mathbf{I}_M+\frac{1}{\sigma^2} \mathbf{H} \mathbf{Q} \mathbf{H}^H\right)\right\},
		\end{aligned}
	\end{equation}
	where $\mathbb{E}_{\widetilde{\mathbf{\Sigma}}}$ denotes the expectation takes over I-CSI. On the other hand, we consider an affine power consumption model \cite{Eeof}. In particular, the total power consumed is denoted as
	\begin{equation}
		P_{\mathrm{tot}}=\omega \operatorname{tr}\left(\mathbf{Q}\right)+N P_{\mathrm{c}}+P_{\mathrm{s}}, 
	\end{equation}
	where $\omega$ describes the transmit power amplifier efficiency, $\operatorname{tr}\left(\mathbf{Q}\right)$ is the total transmit power, $P_{\mathrm{c}}$ denotes the dynamic power consumption associated with each antenna (e.g., circuit power of corresponding radio frequency (RF) signal processing and antenna movement), and $P_{\mathrm{s}}$ accounts for the static circuit power consumption. Consequently, the optimization problem is formulated as follows
	\begin{equation}
		\label{op: original}
		\begin{aligned}
			\max_{\mathbf{t}_n\in \mathcal{C}_{\mathrm{t}}, \mathbf{r}_{m} \in \mathcal{C}_{\mathrm{r}}, \mathbf{Q}\succeq\mathbf{0}} \quad & \frac{ R\left(\mathbf{t},\mathbf{r},\mathbf{Q}\right) } {P_\mathrm{tot}} \\
			\text { s.t. } \quad\qquad
			&\text{C1: }\left\|\mathbf{t}_i-\mathbf{t}_j\right\| \geq D,  \forall i \neq j,\\
			&\text{C2: }\left\|\mathbf{r}_{i}-\mathbf{r}_{j}\right\| \geq D, \forall i \neq j, \,\,\,\, \\
			&\text{C3: } \operatorname{tr}\left(\mathbf{Q}\right) \leq P_\mathrm{max}, 
		\end{aligned}
	\end{equation}
	where $P_\mathrm{max} \geq 0$ is the given maximum available transmit power.
	Note that problem \eqref{op: original} is challenging to obtain a optimal solution for the following reasons. Firstly, the achievable rate in the objective function is non-convex in terms of the APVs and involves expectation. Secondly, the position constraints C1 and C2 are non-convex. As a result, we focus on the alternative optimization of the transmit and receive variables through an AO algorithm combining the DE and SCA approaches.
	
	\section{Design of MA-Enhanced MIMO System}
	\label{sec: algorithm}
	In this section, we first introduce the deterministic functions of the system EE and average achievable rate w.r.t. the transmit and receive variables, respectively. Then, we propose an AO algorithm to alternatively update the transmit and receive variables to maximize the system EE.
	
	\subsection{Deterministic Equivalent and AO Algorithm}
	While the expectation in objective function can usually be approximately computed via the Monte Carlo (MC) method, designing a transmit covariance matrix still poses high computational complexity. Therefore, we propose the DE approach to establish deterministic objective functions in MA-enhanced MIMO systems \cite{SROS}. In particular, for any given $\mathbf r$ the average achievable rate $R\left(\mathbf{t},\mathbf{r},\mathbf{Q}\right)$ w.r.t. the transmit variables can be reformulated by \cite{FDEf}
	\begin{equation}
		\label{eq: DE Rtk+}
		\begin{aligned}
			\bar{R}_{\mathrm{t}}\left(\mathbf{t}, \mathbf{Q}\right) =& \log \operatorname{det}\left( \mathbf{I}_{N}+ \mathbf{G}\left( \mathbf{t} \right)^H\widetilde{\mathbf{\Gamma}} \mathbf{G} \left( \mathbf{t} \right) \mathbf{Q} \right) \\
			+& \log \operatorname{det}\left(\widetilde{\mathbf{\Phi}}\right) - \operatorname{tr}\left( \sigma^2 \left( \mathbf{I}_M - \widetilde{\mathbf{\Phi}} \right) \widetilde{\mathbf{\Theta}} \right)
		\end{aligned}
	\end{equation}
	with matrix $\widetilde{\mathbf{\Gamma}}$ given by
	\begin{equation}
		\begin{aligned}
			& \widetilde{\mathbf{\Gamma}} = - \eta \left( \widetilde{\mathbf{\Theta}} \right) + \sigma^{-2} \overline{\mathbf{\Sigma}}^H \mathbf{F}\left( \mathbf{r} \right) \left(\widetilde{\mathbf{\Phi}}\right)^{-1} \mathbf{F}\left( \mathbf{r} \right)^H \overline{\mathbf{\Sigma}}.
		\end{aligned}
	\end{equation}
	The other matrices are calculated through the following iterative process:
	\begin{subequations}
		\label{eq: DE process}
		\begin{align}
			& \widetilde{\mathbf{\Phi}} = \mathbf{I}_M - \mathbf{F}\left( \mathbf{r} \right)^H \tilde{\eta} \left( \mathbf{\Theta} \right) \mathbf{F}\left( \mathbf{r} \right),
			\\
			& \mathbf{\Phi} = \mathbf{I}_N - \mathbf{Q}^{\frac{H}{2}} \mathbf{G}\left( 	\mathbf{t} \right)^H  \eta \left( \widetilde{\mathbf{\Theta}} \right) \mathbf{G} \left( \mathbf{t} \right) \mathbf{Q}^{\frac{1}{2}}, \\
			& \mathbf{\Theta} = \left( -\sigma^2 \mathbf{\Phi} -  	\mathbf{Q}^{\frac{H}{2}} \overline{\mathbf{H}}^H \left(\widetilde{\mathbf{\Phi}}\right)^{-1} \overline{\mathbf{H}} \mathbf{Q}^{\frac{1}{2}} \right)^{-1}, \\
			& \widetilde{\mathbf{\Theta}} = \left( -\sigma^2 \widetilde{\mathbf{\Phi}} - \overline{\mathbf{H}} \mathbf{Q}^{\frac{1}{2}} \left(\mathbf{\Phi}\right)^{-1} \mathbf{Q}^{\frac{H}{2}} \overline{\mathbf{H}}^H \right)^{-1},
		\end{align}
	\end{subequations}
	where matrix-valued function $\tilde{\eta}$, $\eta$ is written as
	\begin{equation}
		\begin{aligned}
			\tilde{\eta} \left( \mathbf{\Theta} \right) =& \mathbb{E}_{\widetilde{\mathbf{\Sigma}}} \left\{ \widetilde{\mathbf{\Sigma}} \mathbf{G} \left( \mathbf{t} \right) \mathbf{Q}^{\frac{1}{2}} \mathbf{\Theta} \mathbf{Q}^{\frac{H}{2}} \mathbf{G}\left( \mathbf{t} \right)^H  \widetilde{\mathbf{\Sigma}}^H \right\} \\
			=& \operatorname{Diag} \left\{ \mathbf{M} \operatorname{Diag} \left\{	\mathbf{G} \left( \mathbf{t} \right) \mathbf{Q}^{\frac{1}{2}} \mathbf{\Theta} \mathbf{Q}^{\frac{H}{2}} \mathbf{G}\left( \mathbf{t} \right)^H  \right\} \mathbf{M}^H \right\}, \\
			\eta \left( \widetilde{\mathbf{\Theta}} \right) =& \mathbb{E}_{\widetilde{\mathbf{\Sigma}}} \left\{ \widetilde{\mathbf{\Sigma}}^H \mathbf{F}\left( \mathbf{r} \right) \widetilde{\mathbf{\Theta}} \mathbf{F}\left( \mathbf{r} \right)^H \widetilde{\mathbf{\Sigma}} \right\} \\
			=& \operatorname{Diag} \left\{ \mathbf{M}^H \operatorname{Diag} \left\{ \mathbf{F}\left( \mathbf{r} \right) \widetilde{\mathbf{\Theta}} \mathbf{F}\left( \mathbf{r} \right)^H \right\} \mathbf{M} \right\}.
		\end{aligned}
	\end{equation}
	On the other hand, for given any $\mathbf t, \mathbf Q$, the average achievable rate $R\left(\mathbf{t},\mathbf{r},\mathbf{Q}\right)$ w.r.t. the receive APV can be reformulated by
	\begin{equation}
		\begin{aligned}
			\bar{R}_{\mathrm{r}}\left( \mathbf{r} \right) =& \log \operatorname{det}\left( \mathbf{I}_M + \mathbf{F}\left( \mathbf{r} \right)^H  \mathbf{\Gamma} \mathbf{F} \left( \mathbf{r} \right) \right) \\
			+& \log \operatorname{det}\left(\mathbf{\Phi}\right) - \operatorname{tr}\left( \sigma^2 \left( \mathbf{I}_{N} - \mathbf{\Phi} \right) \mathbf{\Theta} \right)
		\end{aligned}
	\end{equation}
	with matrix
	\begin{equation}
		\begin{aligned}
			& \mathbf{\Gamma} = - \widetilde{\eta} \left( \mathbf{\Theta} \right) + \sigma^{-2} \overline{\mathbf{\Sigma}} \mathbf{G} \left( \mathbf{t} \right) \mathbf{Q}^{\frac{1}{2}} \mathbf{\Phi} \mathbf{Q}^{\frac{H}{2}} \mathbf{G}\left( \mathbf{t} \right)^H  \overline{\mathbf{\Sigma}}^H.
		\end{aligned}
	\end{equation}
	
	Based on the above formulations, we propose AO algorithm to optimize EE. In each iteration, we first update the transmit variables, $\left\{\mathbf{t},\mathbf{Q}\right\}$, to maximize EE:
	\begin{equation}
		\label{op: EE t P}
		\begin{aligned}
			\max_{\mathbf{t}_n\in \mathcal{C}_{\mathrm{t}}, \mathbf{Q}\succeq\mathbf{0}} \quad & \frac{ \bar{R}_{\mathrm{t}}\left(\mathbf{t},\mathbf{Q}\right) } {P_\mathrm{tot}} \quad\text { s.t. } \,\, &\text{C1},\text{C3}.
		\end{aligned}
	\end{equation}
	Subsequently, $\bar{R}_{\mathrm{r}}\left(\mathbf{r}\right)$ is maximized by updating receive APV $\mathbf{r}$ as follows
	\begin{equation}
		\label{op: SR r}
		\begin{aligned}
			\max_{\mathbf{r}_{m} \in \mathcal{C}_{\mathrm{r}}} \quad & \bar{R}_{\mathrm{r}}\left(\mathbf{r}\right) \quad\text { s.t. } \,\, &\text{C2}.
		\end{aligned}
	\end{equation}
	
	\subsection{Transmit Design}
	In this subsection, we treat the covariance matrix as a function w.r.t. the transmit APV, and then there is only one variable $\mathbf t$ in the problem. Specifically, given $\mathbf{t}$, we have the inner-layer problem
	\begin{equation}
		\label{op: EE P}
		\begin{aligned}
			\mathbf{Q}\left(\mathbf{t}\right)\triangleq\arg \max_{\mathbf{Q}\succeq\mathbf{0}} \quad & \frac{ \bar{R}_{\mathrm{t}}\left(\mathbf{t},\mathbf{Q}\right) } {P_\mathrm{tot}} \quad \text { s.t. } \,\,
			& \text{C3},
		\end{aligned}
	\end{equation}
	which is a classical concave-convex fractional programming that can be optimally solved by the Dinkelbach's method \cite{EEiW}. However, the resulting subproblem in each iteration may be still of high computational complexity. Consequently, a more efficient and well-structured iterative algorithm for the problem in \eqref{op: EE P} is developed in the next subsection. Then, problem \eqref{op: EE t P} is obviously equivalent to
	\begin{equation}
		\label{op: EE t}
		\begin{aligned}
			\max_{\mathbf{t}_n\in \mathcal{C}_{\mathrm{t}}} \quad & EE\left(\mathbf{t}\right)\triangleq \frac{ \bar{R}_{\mathrm{t}}\left(\mathbf{t},\mathbf{Q}\left(\mathbf{t}\right)\right) } {P_\mathrm{tot}} \quad\text { s.t. } \,\, &\text{C1}.
		\end{aligned}
	\end{equation}
	
	Now, we focus on problem \eqref{op: EE t}. To construct the convex surrogate for optimizing the transmit APV, we begin with the numerical gradient of optimal $EE\left(\mathbf{t}\right)$. The first-order divided difference of $EE\left(\mathbf{t}\right)$ in $i$-th dimension of $\mathbf{t}$ is given by
	\begin{equation}
		\label{eq: numerical gradient t}
		\begin{aligned}
			\left[ \mathbf{\Delta}_{\mathrm{t}}\left(\mathbf{t}\right) \right]_i = \frac{EE\left(\mathbf{t}+\epsilon_1\mathbf{e}_i\right)- EE\left(\mathbf{t}\right)}{\epsilon_1}, i=1,2,\cdots,2N.
		\end{aligned}
	\end{equation}
	where $\epsilon_1$ represents a small change in $i$-th dimension of transmit APV $\mathbf{t}$ and $\mathbf{e}_i$ denotes the standard unit vector in that dimension. On the other hand, from the Cauchy-Schwarz inequality, the position constraints of the transmit MAs in C1 can be replaced with
	\begin{equation}
		\label{eq: t constraint surrogate}
		\begin{aligned}
			\frac{1}{\left\|\mathbf{t}_i^{\ell,\hat{\ell}_{\mathrm{t}}}-\mathbf{t}_j^{\ell,\hat{\ell}_{\mathrm{t}}}\right\|_2}\left(\mathbf{t}_i-\mathbf{t}_j\right)^T\left(\mathbf{t}_i^{\ell,\hat{\ell}_{\mathrm{t}}}-\mathbf{t}_j^{\ell,\hat{\ell}_{\mathrm{t}}}\right) \geq D,\quad i \neq j,
		\end{aligned}
	\end{equation}
	in iteration $\hat{\ell}_{\mathrm{t}}+1$ of the proposed SCA algorithm for subproblem \eqref{op: EE t} in iteration $\ell+1$ of the AO algorithm.
	Then, the convex surrogate of problem \eqref{op: EE t} can be given by
	\begin{equation}
		\label{op: SCA t surrogate}
		\begin{aligned}
			\max_{\mathbf{t}_n\in \mathcal{C}_{\mathrm{t}}} & -\delta_{\mathrm{t}}(\mathbf{t}-\mathbf{t}^{\ell,\hat{\ell}_{\mathrm{t}}})^T(\mathbf{t}-\mathbf{t}^{\ell,\hat{\ell}_{\mathrm{t}}}) + \mathbf{\Delta}_{\mathrm{t}}(\mathbf{t}^{\ell,\hat{\ell}_{\mathrm{t}}})^T (\mathbf{t}-\mathbf{t}^{\ell,\hat{\ell}_{\mathrm{t}}}) \\
			\text { s.t. } \quad
			&\eqref{eq: t constraint surrogate},
		\end{aligned}
	\end{equation}
	where $\delta_{\mathrm{t}}>0$ guarantees the strong concavity of the objective. To reduce computational complexity, we consider the problem similar to that in \eqref{op: SCA t surrogate} but without constraint \eqref{eq: t constraint surrogate}, whose solution is trivially 
	$\tilde{\mathbf{t}}^{\ell,\hat{\ell}_{\mathrm{t}}+1}=\left[ \frac{\mathbf{\Delta}_{\mathrm{t}}\left(\mathbf{t}^{\ell,\hat{\ell}_{\mathrm{t}}}\right)}{2\delta_t} \right]_{\mathcal{C}_{\mathrm{t}}}$.
	Then, if $\tilde{\mathbf{t}}^{\ell,\hat{\ell}_{\mathrm{t}}+1}$ satisfies constraint \eqref{eq: t constraint surrogate}, it is also the optimal solution to problem \eqref{op: SCA t surrogate}. Otherwise, problem \eqref{op: SCA t surrogate} can be efficiently solved by traditional convex optimization techniques, such as CVX. The transmit APV is updated according to
	\begin{equation}
		\label{eq: update t}
		\begin{aligned}
			\mathbf{t}^{\ell,\hat{\ell}_{\mathrm{t}}+1}=\left(1-\tau_{\mathrm{t}}^{\ell,\hat{\ell}_{\mathrm{t}}+1}\right)\mathbf{t}^{\ell,\hat{\ell}_{\mathrm{t}}}+\tau_{\mathrm{t}}^{\ell,\hat{\ell}_{\mathrm{t}}+1}\bar{\mathbf{t}}^{\ell,\hat{\ell}_{\mathrm{t}}+1},
		\end{aligned}
	\end{equation}
	where $\bar{\mathbf{t}}^{\ell,\hat{\ell}_{\mathrm{t}}+1}$ and $\tau_{\mathrm{t}}^{\ell,\hat{\ell}_{\mathrm{t}}+1}$ are the solution to problem \eqref{op: SCA t surrogate} and the step size selected by backtracking line search, respectively. In each iteration, we start with a large positive step size, $\tau_{\mathrm{t}}^{\ell,\hat{\ell}_{\mathrm{t}}+1}=\tau^0$, and repeatedly reduce it to $\tau\tau_{\mathrm{t}}^{\ell,\hat{\ell}_{\mathrm{t}}+1}$ with a factor $\tau\in(0,1)$, until the following Armijo–Goldstein condition is satisfied:
	\begin{equation}
		\label{eq: Armijo–Goldstein condition t}
		\begin{aligned}
			I_{\mathrm{t}}\left(\mathbf{t}^{\ell,\hat{\ell}_{\mathrm{t}}+1},\mathbf{t}^{\ell,\hat{\ell}_{\mathrm{t}}}\right) \geq \xi \tau_{\mathrm{t}}^{\ell,\hat{\ell}_{\mathrm{t}}+1} \left\|	\bar{\mathbf{t}}^{\ell,\hat{\ell}_{\mathrm{t}}+1}-\mathbf{t}^{\ell,\hat{\ell}_{\mathrm{t}}} \right\|^2,
		\end{aligned}
	\end{equation}
	where $I_{\mathrm{t}}\left(\mathbf{t}^{\ell,\hat{\ell}_{\mathrm{t}}+1},\mathbf{t}^{\ell,\hat{\ell}_{\mathrm{t}}}\right) \triangleq EE\left(\mathbf{t}^{\ell,\hat{\ell}_{\mathrm{t}}+1}\right) - EE\left(\mathbf{t}^{\ell,\hat{\ell}_{\mathrm{t}}}\right)$ is the increment of the objective in problem \eqref{op: EE t} and $\xi \in (0, 1)$ is a given control parameter to guarantee that  the objective function achieves an adequate increase. The SCA algorithm is terminated when $\hat{\ell}_{\mathrm{t}} = \hat{L}_{\mathrm{t}}$ or the increment is less than $\epsilon_2$, i.e.,
	\begin{equation}
		\label{eq: termin condition t}
		\begin{aligned}
			I_{\mathrm{t}}\left(\mathbf{t}^{\ell,\hat{\ell}_{\mathrm{t}}+1},\mathbf{t}^{\ell,\hat{\ell}_{\mathrm{t}}}\right) \leq \epsilon_2.
		\end{aligned}
	\end{equation}

	\subsection{Inner-Layer Problem Design}
	To solve the problem \eqref{op: EE P}, we derive the optimal solution for the unconstrained EE maximization problem:
	\begin{equation}
		\label{op: Unconstrained EE P}
		\begin{aligned}
			\mathbf{Q}_{\mathrm{opt}}^{\ell}=\arg \max_{\mathbf{Q}\succeq\mathbf{0}} \quad & \frac{ \bar{R}_{\mathrm{t}}\left(\mathbf{t},\mathbf{Q}\right) } {P_\mathrm{tot}}.
		\end{aligned}
	\end{equation}
	Note that the optimal transmit covariance matrix for the problem \eqref{op: EE P} is equal to $\mathbf{Q}_{\mathrm{opt}}^{\ell}$ if the corresponding power consumption satisfies the constraint, i.e., $\operatorname{tr}\left(\mathbf{Q}_{\mathrm{opt}}^{\ell}\right)\leq P_\mathrm{max}$. Otherwise, the maximum EE is achieved while transmitting with full power budget $P_\mathrm{max}$. Thus, the sum rate maximization problem is then considered as follows
	\begin{equation}
		\label{op: SR P}
		\begin{aligned}
			\widetilde{\mathbf{Q}}_{\mathrm{opt}}^{\ell}=\arg \max_{\mathbf{Q}\succeq\mathbf{0}} \quad & \bar{R}_{\mathrm{t}}\left(\mathbf{t},\mathbf{Q}\right)\quad \text { s.t. } 
			& \operatorname{tr}\left(\mathbf{Q}\right) = P_\mathrm{max}.
		\end{aligned}
	\end{equation}
	Subsequently, the above two problems \eqref{op: Unconstrained EE P} and \eqref{op: SR P} are optimally tackled by the Dinkelbach’s method and the water-filling algorithm, respectively. In particular, for the former, we iteratively solve the quadratic-form optimization problem, which is written as
	\begin{equation}
		\label{op: Dinkelbach P}
		\begin{aligned}
			\max_{\mathbf{Q}\succeq\mathbf{0}} \quad & \bar{R}_{\mathrm{t}}\left(\mathbf{t},\mathbf{Q}\right) - \eta^{\ell,\hat{\ell}_{\mathrm{q}}}P_\mathrm{tot}
		\end{aligned}
	\end{equation}
	where
	\begin{equation}
		\label{op: Dinkelbach eta}
		\begin{aligned}
			\eta^{\ell,\hat{\ell}_{\mathrm{q}}} &= \left.\frac{ \bar{R}_{\mathrm{t}}\left(\mathbf{t},\mathbf{Q}\right) } {P_\mathrm{tot}} \right|_{\mathbf{Q}=\mathbf{Q}^{\ell,\hat{\ell}_{\mathrm{q}}}}.
		\end{aligned}
	\end{equation}
	The closed-form solution to problem \eqref{op: Dinkelbach P} in iteration $\hat{\ell}_{\mathrm{q}}+1$ of Dinkelbach’s method satisfies
	\begin{equation}
		\label{eq: Unconstrained EE optimal P}
		\begin{aligned}
			\mathbf{Q}^{\ell,\hat{\ell}_{\mathrm{q}}+1}=\mathbf{U}_{\mathrm{G}}\left[ \left(\eta^{\ell,\hat{\ell}_{\mathrm{q}}}\right)^{-1}\mathbf{I}_N -  \mathbf{\Lambda}_{\mathrm{G}}^{\dagger} \right]^+\mathbf{U}_{\mathrm{G}}^H,
		\end{aligned}
	\end{equation}
	where $\mathbf{U}_{\mathrm{G}}$  is the square $N\times N$ matrix whose each column is the eigenvector of matrix $\mathbf{G} \left( \mathbf{t} \right)^H  \widetilde{\mathbf{\Gamma}} \mathbf{G} \left( \mathbf{t} \right)$ and $\mathbf{\Lambda}_{\mathrm{G}}$ is the diagonal matrix whose main diagonal elements are the corresponding eigenvalues. On the other hand, the optimal transmit covariance matrix for problem \eqref{op: SR P} follow the classical water-filling structure as follows
	\begin{equation}
		\begin{aligned}
			\widetilde{\mathbf{Q}}_{\mathrm{opt}}^{\ell}=\mathbf{U}_{\mathrm{G}}\left[ \mu^{-1}\mathbf{I}_N -  \mathbf{\Lambda}_{\mathrm{G}}^{\dagger} \right]^+\mathbf{U}_{\mathrm{G}}^H
		\end{aligned}
	\end{equation}
	with $\mu$ satisfying $\sum_{i=1}^N\left[ \left[ \mu^{-1}\mathbf{I}_N -  \mathbf{\Lambda}_{\mathrm{G}}^{\dagger} \right]^+ \right]_{i,i}=P_{\mathrm{max}}$, which can be found by the bisection method.
	
	\subsection{Receive Design}
	To solve problem \eqref{op: SR r} efficiently, we propose a low-complexity SCA algorithm, where the structure of the objective is leveraged to construct a convex surrogate problem. In particular, we define matrix
	\begin{equation}
		\begin{aligned}
			\hat{\mathbf{J}}\left(\mathbf{r}\right)&\triangleq\mathbf{\Delta}_{X_{\mathrm{r}}}\operatorname{Diag}(\mathbf{x}_{\mathrm{r}}-\mathbf{x}_{\mathrm{r}}^{\ell,\hat{\ell}_{\mathrm{r}}})+\operatorname{Diag}(\mathbf{x}_{\mathrm{r}}-\mathbf{x}_{\mathrm{r}}^{\ell,\hat{\ell}_{\mathrm{r}}})\mathbf{\Delta}_{X_{\mathrm{r}}}^H \\
			&+\mathbf{\Delta}_{Y_{\mathrm{r}}}\operatorname{Diag}(\mathbf{y}_{\mathrm{r}}-\mathbf{y}_{\mathrm{r}}^{\ell,\hat{\ell}_{\mathrm{r}}})+\operatorname{Diag}(\mathbf{y}_{\mathrm{r}}-\mathbf{y}_{\mathrm{r}}^{\ell,\hat{\ell}_{\mathrm{r}}})\mathbf{\Delta}_{Y_{\mathrm{r}}}^H,
		\end{aligned}
	\end{equation}
	where 
	\begin{equation}
		\label{eq: Delta XYr}
		\begin{aligned}
			\mathbf{x}_{\mathrm{r}} \triangleq& \left(x_{\mathrm{r},1},\cdots,x_{\mathrm{r},M}\right)^T,
			\mathbf{y}_{\mathrm{r}} \triangleq \left(y_{\mathrm{r},1},\cdots,y_{\mathrm{r},M}\right)^T,\\
			\mathbf{\Delta}_{X_{\mathrm{r}}} =& \frac{j2\pi}{\lambda} \mathbf{F}\left( \mathbf{r} \right)^H  \mathbf{\Gamma} \\
			&\operatorname{Diag}\left( \sin\theta_{\mathrm{r}}^{1}\cos\phi_{\mathrm{r}}^{1},\cdots,\sin\theta_{\mathrm{r}}^{L_{\mathrm{r}}}\cos\phi_{\mathrm{r}}^{L_{\mathrm{r}}} \right)^T \mathbf{F} \left( \mathbf{r} \right), \\
			\mathbf{\Delta}_{Y_{\mathrm{r}}} =& \frac{j2\pi}{\lambda} \mathbf{F}\left( \mathbf{r} \right)^H  \mathbf{\Gamma} \operatorname{Diag} \left(
			\cos\theta_{\mathrm{r}}^{1},\cdots,\cos\theta_{\mathrm{r}}^{L_{\mathrm{r}}} \right)^T \mathbf{F} \left( \mathbf{r} \right).
		\end{aligned}
	\end{equation}
	Then, the concave surrogate of $\bar{R}_{\mathrm{r}}\left(\mathbf{r}\right)$ can be constructed as quadratic-form function
	\begin{equation}
		\begin{aligned}
			\hat{R}_{\mathrm{r}}\left(\mathbf{r}\right) &=-\operatorname{Tr} \left\{ \mathbf{E} \hat{\mathbf{J}}\left(\mathbf{r}\right) \mathbf{E} \hat{\mathbf{J}}\left(\mathbf{r}\right) \right\}+\operatorname{Tr} \left\{ \mathbf{E} \hat{\mathbf{J}}\left(\mathbf{r}\right) \right\} \\
			&- \delta_{\mathrm{r}} \left( \mathbf{r}-\mathbf{r}^{\ell,\hat{\ell}_{\mathrm{r}}} \right)^T \left( \mathbf{r}-\mathbf{r}^{\ell,\hat{\ell}_{\mathrm{r}}} \right)+ \bar{R}_{\mathrm{r}}\left(\mathbf{r}^{\ell,\hat{\ell}_{\mathrm{r}}}\right).
		\end{aligned}
	\end{equation}
	where $\delta_{\mathrm{r}}>0$ ensures that the objective is strongly concave w.r.t. $\mathbf{r}$ and $\mathbf{E}=\left( \mathbf{I}_M + \mathbf{F}\left( \mathbf{r} \right)^H  \mathbf{\Gamma} \mathbf{F} \left( \mathbf{r} \right) \right)^{-1}$. 
	Moreover, the position constraints of the receive MAs in C2 is replaced with
	\begin{equation}
		\label{eq: r constraint surrogate}
		\begin{aligned}
			\frac{1}{\left\|\mathbf{r}_{i}^{\ell,\hat{\ell}_{\mathrm{r}}}-\mathbf{r}_{j}^{\ell,\hat{\ell}_{\mathrm{r}}}\right\|_2}\left(\mathbf{r}_{i}-\mathbf{r}_{j}\right)^T\left(\mathbf{r}_{i}^{\ell,\hat{\ell}_{\mathrm{r}}}-\mathbf{r}_{j}^{\ell,\hat{\ell}_{\mathrm{r}}}\right) \geq D,\quad i \neq j.
		\end{aligned}
	\end{equation}
	
	Based on the above discussion, we solve the following convex surrogate problem in iteration $\hat{\ell}_{\mathrm{r}}+1$ of our proposed SCA algorithm as shown in
	\begin{equation}
		\label{op: SCA r surrogate}
		\begin{aligned}
			\max_{\mathbf{r}_{m} \in \mathcal{C}_{\mathrm{r}}} \quad & \hat{R}_{\mathrm{r}}\left(\mathbf{r}\right) \quad \text { s.t. } \,\,
			\eqref{eq: r constraint surrogate}.
		\end{aligned}
	\end{equation}To further reduce computational complexity, we consider a problem similar with \eqref{op: SCA r surrogate} but without the constraints, i.e., $\max_{\mathbf{r}} \, \hat{R}_{\mathrm{r}}\left(\mathbf{r}\right)$. The above problem is optimally solved when the first-order optimality condition is satisfied, which is in fact solving a system of linear equations. After that, if the solution satisfies constraints $\mathbf{r}_{m} \in \mathcal{C}_{\mathrm{r}}$ and \eqref{eq: r constraint surrogate}, it is the optimal solution to problem \eqref{op: SCA r surrogate}. Otherwise, problem \eqref{op: SCA r surrogate} can be efficiently solved by traditional convex optimization techniques, such as CVX. The update of the receive APV in each iteration and the termination condition of the SCA algorithm is similar to that for the transmit design. 
	
	\subsection{Complexity and Convergence Analysis}
	The overall AO algorithm is summarized in Algorithm \ref{Al: AO}. In Step 1, we initialize $\mathbf{t}^0$, $\mathbf{r}^0$ and $\mathbf{Q}^0$ satisfying all constraints and then they are optimized in a alternative style in Step 2-7.
	\begin{algorithm}[h]
		\caption{AO Algorithm for Solving Problem \eqref{op: original}}
		\label{Al: AO}
		\begin{spacing}{1}
			\textbf{Input:} $N$, $M$, $\sigma^2$, $\lambda$, $D$, $\mathcal{C}_{\mathrm{t}}$, $\mathcal{C}_{\mathrm{r}}$, $\omega$, $P_{\mathrm{max}}$, $P_{\mathrm{c}}$, $P_{\mathrm{s}}$; $\epsilon_1$, $\epsilon_2$, $L_{\mathrm{ao}}$, $\hat{L}_{\mathrm{t}}$, $\hat{L}_{\mathrm{r}}$; S-CSI \\
			\textbf{Output:} $\mathbf{t}^{L_{\mathrm{ao}}}$, $\mathbf{r}^{L_{\mathrm{ao}}}$ and $\mathbf{Q}^{L_{\mathrm{ao}}}$
			\begin{algorithmic}[1]
				\STATE{Initialize $\mathbf{t}^0$, $\mathbf{r}^0$ and $\mathbf{Q}^0$, satisfying all constraints}
				\FOR{$\ell = 1 \text{ to } L_{\mathrm{ao}}$}
				
				\STATE{Construct $\bar{R}_{\mathrm{t}}\left(\mathbf{t},\mathbf{Q}\right)$ by iterative process \eqref{eq: DE process}}
				
				\STATE{Calculate $\mathbf{t}^{\ell}$ and $\mathbf{Q}^{\ell}$ by solving problem \eqref{op: EE t P}}
				
				\STATE{Construct $\bar{R}_{\mathrm{r}}\left(\mathbf{r}\right)$ by iterative process \eqref{eq: DE process}}
				
				\STATE{Calculate $\mathbf{r}^{\ell}$ by solving problem \eqref{op: SR r}}
				
				\ENDFOR
				\STATE{\textbf{return} $\mathbf{t}^{L_{\mathrm{ao}}}$, $\mathbf{r}^{L_{\mathrm{ao}}}$ and $\mathbf{Q}^{L_{\mathrm{ao}}}$}
			\end{algorithmic}
		\end{spacing}
	\end{algorithm}
	
	The convergence of Algorithm \ref{Al: AO} is analyzed as follows. 
	It is apparent that the proposed AO algorithm generates a non-decreasing sequence $\left\{\mathbf{t}^{\ell}, \mathbf{r}^{\ell}, \mathbf{Q}^{\ell}\right\}_{\ell=1}^{\infty}$. Besides, the objective of problem \eqref{op: original} is upper-bounded by a finite value since the feasible region is compact. Thus, the objective value convergence of Algorithm \ref{Al: AO} is guaranteed. 
	On the other hand, the computational complexity of a DE iterative process, the SCA algorithms for solving problems \eqref{op: EE t P} and \eqref{op: SR r} are $\mathcal{O}\left(L_{\mathrm{de}}N^{3}+L_{\mathrm{de}}M^{3}\right)$, $\mathcal{O}\left(\hat{L}_{\mathrm{t}}N^4+\hat{L}_{\mathrm{t}}N^{3.5}\log(1/\epsilon)\right)$ and
	$\mathcal{O}\left(\hat{L}_{\mathrm{r}}M^{3.5}\log(1/\epsilon)\right)$, respectively, where $L_{\mathrm{de}}$, $L_{\mathrm{t}}$ and $L_{\mathrm{r}}$ are the number of iterations. Consequently, the computational complexity of Algorithm \ref{Al: AO} is $\mathcal{O}\left( \left( \hat{L}_{\mathrm{t}}N^4+\hat{L}_{\mathrm{t}}N^{3.5}\log(1/\epsilon)+\hat{L}_{\mathrm{r}}M^{3.5}\log(1/\epsilon)+L_{\mathrm{de}}N^{3}\right.\right.$$\\\left.\left.+L_{\mathrm{de}}M^{3} \right) L_{\mathrm{ao}} \right)$.
	
	\section{Numerical Results}
	\label{sec: numerical results}
	In this section, we numerically evaluate the performance of the proposed single-user MA-enhanced MIMO system. In our simulations,  the distance between Tx and Rx, $d$, follows a uniform distribution between $20$ and $100$ meters (m). The channel gain is set as $g = c_0 d^{-\alpha_0}$, where $c_0$ denotes the expected value of the path loss at the reference distance of $1$ m, and $\alpha_0$ represents the path loss exponent. Moreover, we assume that there are $L_{\mathrm{t}}=L_{\mathrm{r}}=L$ paths and the entries of $\overline{\mathbf{\Sigma}}$ are set as $0$ except $\left[\overline{\mathbf{\Sigma}}\right]_{1,1} = \sqrt{ \frac{g K_\mathrm{r}}{\left(K_\mathrm{r}+1\right)} }$ as well as the entries of $\mathbf{M}$ are set as $0$ except $\left[\mathbf{M}\right]_{l,l} = \sqrt{ \frac{g}{\left(L-1\right)\left(K_\mathrm{r}+1\right)} }$, $2 \leq l \leq L$, where $K_\mathrm{r}$ is the Rician factor. Moreover, the $L$ pairs of elevation/azimuth AoDs and AoAs are i.i.d. random variables following distributions $f\left(\theta_{\mathrm{t}}^l,\phi_{\mathrm{t}}^l\right)=\frac{1}{2\pi} \sin \phi_{\mathrm{t}}^l$, $\theta_{\mathrm{t}}^l\in\left[0,\pi\right]$, $\phi_{\mathrm{t}}^l\in\left[0,\pi\right]$ and $f\left(\theta_{\mathrm{r}}^l,\phi_{\mathrm{r}}^l\right)=\frac{1}{2\pi} \sin \phi_{\mathrm{r}}^l$, $\theta_{\mathrm{r}}^l\in\left[0,\pi\right]$, $\phi_{\mathrm{r}}^l\in\left[0,\pi\right]$, $1\leq l \leq L$, respectively. The size of the transmit and receive regions are set as $X\times X$.
		\begin{table}[htb]
		\begin{center}
			\caption{Simulation Parameters.}
			\label{Tb: simulation parameter}
			\begin{tabular}{|c|c|c|c|}
				\hline Parameter & Value & Parameter & \,\, Value \,\, \\
				\hline $N$, $M$ & $4$ & $P_{\mathrm{max}}$, $P_{\mathrm{c}}$ & $30\mathrm{~dBm}$ \\
				\hline $D$ & $0.5\lambda$ & $P_{\mathrm{s}}$ & $40\mathrm{~dBm}$ \\
				\hline $X$ & $2\lambda$ & $L_{\mathrm{ao}}$, $L_{\mathrm{de}}$ & $20$ \\
				\hline $L$ & $5$ & $\hat{L}_{\mathrm{t}}$, $\hat{L}_{\mathrm{r}}$ & 20 \\
				
				\hline $K_\mathrm{r}$ & $10$ & $\delta_{\mathrm{t}}$, $\delta_{\mathrm{r}}$ & $0.02$ \\
				\hline $\sigma^2$ & $-80\mathrm{~dBm}$ & $\tau^0$ & $1$ \\
				\hline $c_0$ & $-40\mathrm{~dB}$ & $\tau$ & $0.5$ \\
				\hline $\alpha_0$ & $2.8$ & $\xi$ & $0.6$ \\
				\hline $\omega$ & $5$ & $\epsilon_1$, $\epsilon_2$ & $10^{-3}$ \\
				\hline 
			\end{tabular}
		\end{center}
	\end{table}
	The default settings for the simulation parameters are provided in Table \ref{Tb: simulation parameter} \cite{Mccf}.
	We compare the performance of the proposed \textbf{MA}-enhanced system with the following schemes, where the antennas in the fixed-position UPAs are spaced by $0.5\lambda$:
	\begin{itemize}
		\item [$\bullet$]\textbf{TMA}: The MAs are only deployed at Tx and Rx is equipped with $2\times2$ UPA.
		\item [$\bullet$]\textbf{RMA}: The MAs are only deployed at Rx and Tx is equipped with $2\times2$ UPA.
		\item [$\bullet$]\textbf{UPA}: Tx and Rx are both equipped with $2\times2$ UPAs.
	\end{itemize}

	\subsection{Impact of the Maximum Transmit Power}
	\begin{figure}[!h]
		\centering
		\includegraphics[width=0.4\textwidth,height=0.3\textwidth]{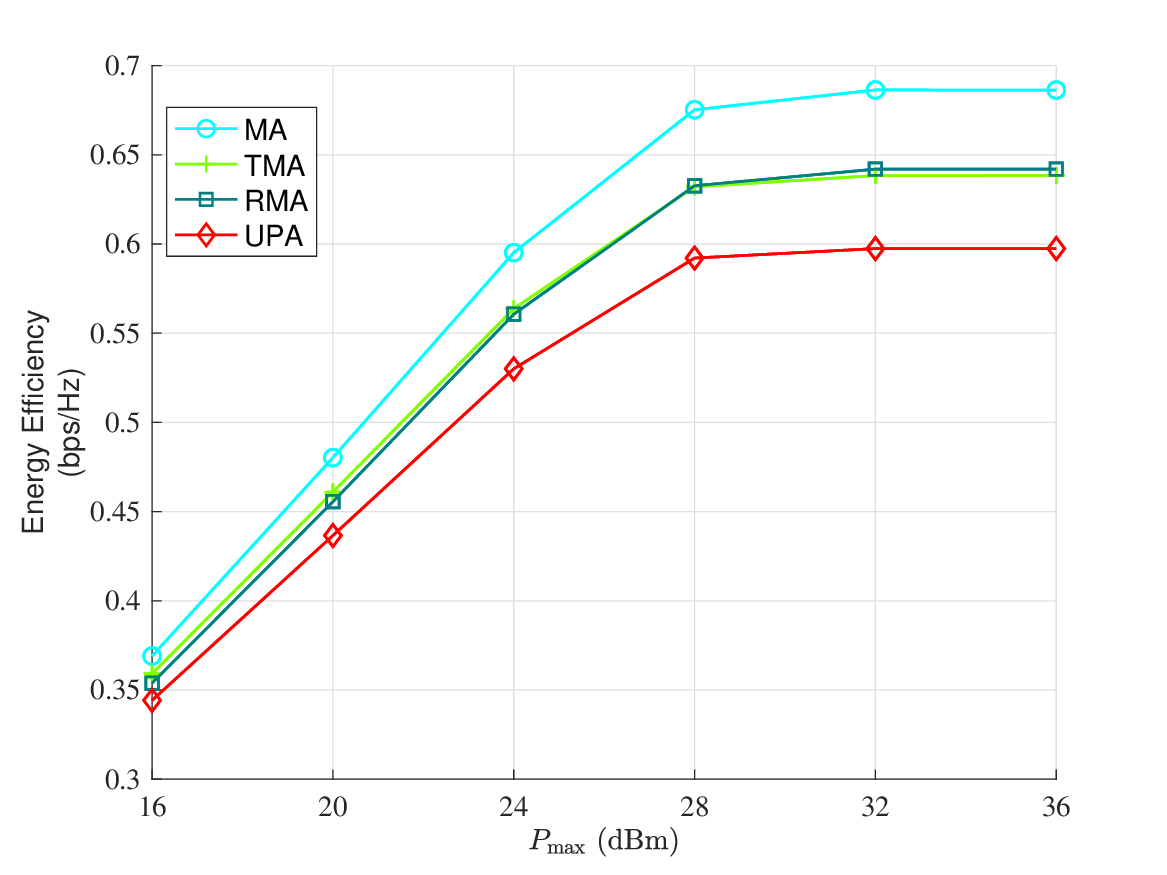}
		\caption{EE versus the maximum transmit power.}
		\label{fig: Simulation P}
	\end{figure}
	From Fig. \ref{fig: Simulation P}, it can be seen that the system EE increases with the maximum transmit power and finally saturates when $P_{\mathrm{max}}$ is greater than $32 \mathrm{~dBm}$. This is because the enhancement in the system rate, resulting from the additional transmit power, is counterbalanced by the rise in the denominator when considering EE. Consequently, excess transmit power that exceeds the optimal power threshold cannot introduce any improvement in the EE performance. Moreover, due to different amounts of spatial degree of freedoms (DoFs) available for the schemes under comparison, the results demonstrate that the TMA and RMA schemes outperform the UPA scheme. Notably, our proposed MA system achieves the best EE performance across all schemes for any power level.
	
	\subsection{Impact of the Movement Region Size}
	\begin{figure}
		\centering
		\includegraphics[width=0.4\textwidth,height=0.3\textwidth]{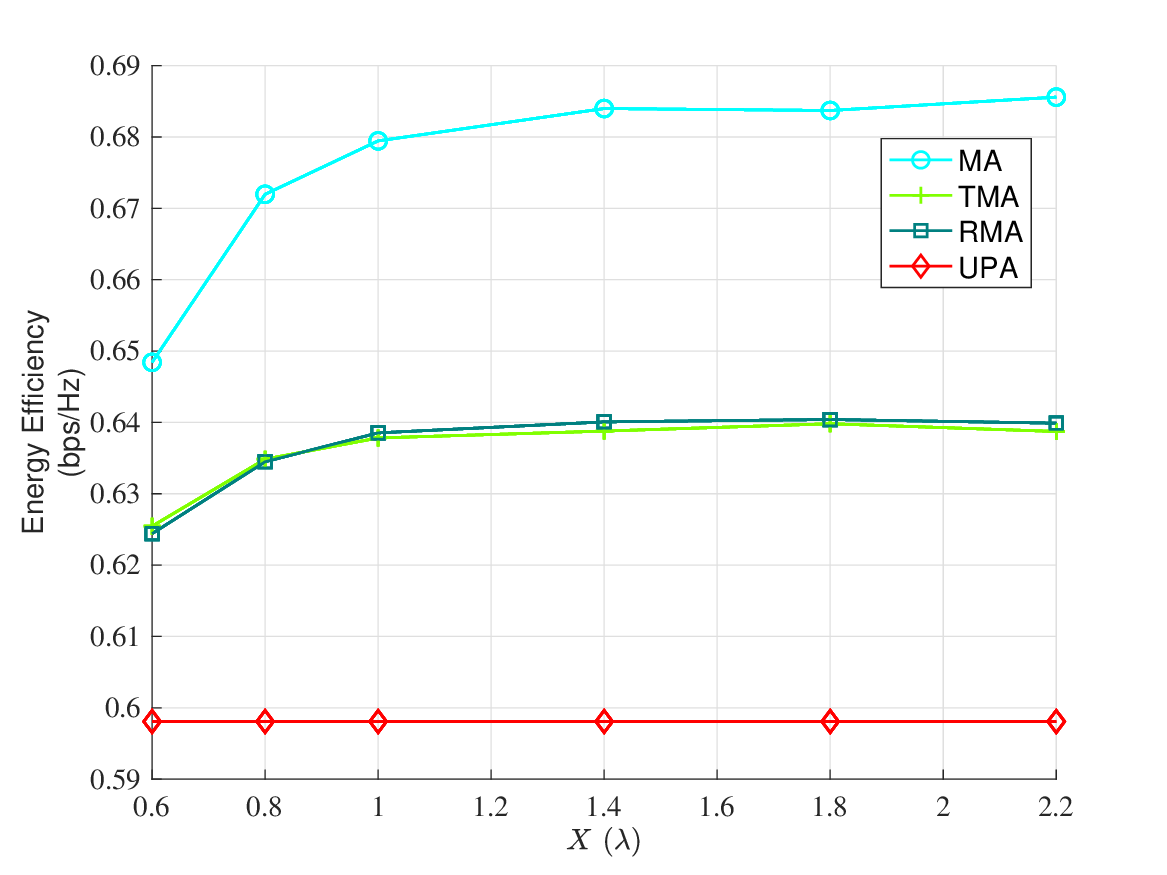}
		\caption{EE versus the size of the movement region.}
		\label{fig: Simulation T}
	\end{figure}
	
	Fig. \ref{fig: Simulation T} shows the impact of the region size on the EE performance. Since the increase of $X$ brings about larger movement regions for the MA design, the EE of the MA, TMA and RMA schemes rises with $X$. Furthermore, all schemes converge when the region size surpasses $1.4\lambda$, suggesting that a nearly optimal system EE can be attained with a finite size. With $X=2.2\lambda$, the proposed scheme achieves $7.3\%$, $7.1\%$ and $14.6\%$ EE improvements over the TMA, RMA and UPA schemes, respectively.
	
	\section{Conclusion}
	\label{sec: conclusion}
	In this paper, we investigated a single-user MIMO system with S-CSI, where Tx and Rx are equipped with multiple MAs. The system EE was maximized by optimizing the transmit covariance matrix and the APVs through an AO algorithm. Firstly, we introduced the deterministic functions of the system EE and the average achievable rate w.r.t. the transmit variables and the receive APV, respectively. Subsequently, two SCA algorithms were developed to alternatively optimize the transmit variables and the receive APV based on the rule of maximizing the above deterministic functions of the system EE and the average achievable rate,respectively. Finally, numerical results revealed that the proposed MA-enhanced systems can significantly improve the system EE compared to several benchmark schemes and the optimal performance can be achieved with a finite size of movement regions for MAs.
	
	\begin{appendices}
		
	\end{appendices}

\end{document}